\documentclass{article}
\usepackage{color}
\usepackage{graphicx}
\usepackage{amsmath, amsthm}
\usepackage{mathtools}
\usepackage{natbib}
\usepackage{url}
\usepackage{soulutf8}
\RequirePackage[colorlinks,citecolor=blue,urlcolor=blue]{hyperref}
\usepackage{xcolor}
\usepackage{chngcntr}
\usepackage{subfig}
\usepackage{hyperref}
\usepackage{algorithm}
\usepackage[noend]{algpseudocode}
\usepackage{setspace}
\newtheorem{theorem}{Theorem}

\usepackage{xfrac}
\definecolor{forest}{rgb}{0.133,0.545,0.133}
\usepackage{multirow}
\usepackage{amsfonts}

\evensidemargin=\oddsidemargin
\usepackage{etoolbox}

\newif\ifabbreviation
\pretocmd{\thebibliography}{\abbreviationfalse}{}{}
\AtBeginDocument{\abbreviationtrue}

\begin{document}
	\newcommand{\bb}{\boldsymbol{\beta}}

	\title{An Economical Approach to Design Posterior Analyses}



	\author{Luke Hagar\footnote{Luke Hagar is the corresponding author and may be contacted at \url{luke.hagar@mail.mcgill.ca}.} \hspace{35pt} Nathaniel T. Stevens$^{\dagger}$ \bigskip \\ 
 $^*$\textit{Department of Epidemiology, Biostatistics \& Occupational Health, McGill University} \\ $^{\dagger}$\textit{Department of Statistics \& Actuarial Science, University of Waterloo}}

	\date{}

	\maketitle

	\begin{abstract}

To design Bayesian studies, criteria for the operating characteristics of posterior analyses  – such as power and the type I error rate – are often assessed by estimating sampling distributions of posterior probabilities via simulation. In this paper, we propose an economical method to determine optimal sample sizes and decision criteria for such studies. Using our theoretical results that model posterior probabilities as a function of the sample size, we assess operating characteristics throughout the sample size space given simulations conducted at only two sample sizes. These theoretical results are used to construct bootstrap confidence intervals for the optimal sample sizes and decision criteria that reflect the stochastic nature of simulation-based design. We also repurpose the simulations conducted in our approach to efficiently investigate various sample sizes and decision criteria using contour plots. The broad applicability and wide impact of our methodology is illustrated using two clinical examples.

		\bigskip

	\end{abstract}

	\maketitle

	\baselineskip=19.5pt


  \section{Introduction}\label{sec:intro.new}

   In Bayesian methods for data-driven decision making, a scalar estimand often quantifies the impact of two choices available to decision makers. Such decisions are regularly informed by assessing a hypothesis $H_1$ based on the posterior distribution of this estimand. Decision-making methods with posterior probabilities have been proposed in various settings (see e.g., \citet{berry2010bayesian, brutti2014bayesian, stevens2022comparative}). The observed data provide sufficient evidence to support $H_1$ if the posterior probability $Pr(H_1 ~|~ data)$ is greater than or equal to a critical value $\gamma \in [0.5, 1)$. When comparing complementary hypotheses $H_1$ and $H_0$, decision-making methods with Bayes factors \citep{jeffreys1935some, kass1995bayes, morey2011bayes} can be viewed as a special case of those with posterior probabilities \citep{hagar2023fast}. This paper therefore focuses on posterior probabilities, though the methods extend to the use of Bayes factors. 
   
   It is important that decision-making methods with posterior summaries yield trustworthy conclusions. In clinical trials, regulatory agencies require that Bayesian designs are assessed with respect to frequentist operating characteristics \citep{fda2019adaptive}. These design procedures are called hybrid approaches to sample size determination since they leverage theory from Bayesian and frequentist statistics \citep{berry2010bayesian}. Decision makers in nonclinical settings may also want to control the power and type I error rate of Bayesian designs (see e.g., \citet{deng2015objective} and \citet{larsen2024statistical} for a discussion of operating characteristics in online A/B tests). Power is the probability that $Pr(H_1 ~|~ data) \ge \gamma$ according to some data generation process where $H_1$ is true. Under this data generation process, the study power should be at least $1 - \beta$. The type I error rate is the probability that $Pr(H_1 ~|~ data) \ge \gamma$ according to some data generation process where $H_1$ is false (and $H_0$ is true). Given a data generation process under $H_0$, the type I error rate for the study should not exceed $\alpha$.

Design inputs for Bayesian studies include the hypotheses $H_1$ and $H_0$, data generation processes, prior distributions, treatment allocations, and values for $\alpha$ and $\beta$. For a particular set of design inputs, the sample size $n$ and critical value $\gamma$ determine whether criteria for the operating characteristics of the posterior analysis are satisfied. To support flexible study design, $(n, \gamma)$ combinations that control these operating characteristics can be found using intensive simulation \citep{wang2002simulation}. In general, many samples of size $n$ are simulated according to $H_1$ ($H_0$) to estimate the sampling distribution of posterior probabilities and the proportion of samples for which $Pr(H_1 ~|~ data) \ge \gamma$ estimates power (the type I error rate). This process is repeated for various sample sizes using various critical values until a suitable $(n, \gamma)$ combination is found, and this computational burden is compounded over all combinations of the design inputs that practitioners wish to investigate. Moreover, the impact of simulation variability on the recommended $(n, \gamma)$ combination is underreported since quantifying this impact often involves conducting further numerical studies \citep{wilson2021efficient}. An economical framework to determine the $(n, \gamma)$ combination that minimizes the sample size $n$ while satisfying criteria for both operating characteristics would meaningfully expedite the design of posterior analyses.

   Recently, several strategies have been employed to reduce the computational burden associated with controlling operating characteristics for Bayesian studies. Certain strategies are tailored to specific statistical distributions (see e.g., \citet{shi2019control} for the efficient design of sequential Bernoulli trials). Other approaches accommodate a variety of statistical models. One such general strategy leverages the parametric efficiency of beta distributions fit to the sampling distribution of posterior probabilities \citep{golchi2022estimating, golchi2023estimating}. Another general strategy prioritizes exploring segments of the sampling distribution of posterior probabilities such that $Pr(H_1 \hspace{1pt}| \hspace{1pt}data) \approx \gamma$ \citep{hagar2023fast}. The use of sampling distribution segments is efficient, but those methods are difficult to implement in complex design scenarios. Here, we present a remarkably simple method to design posterior analyses that does not impose parametric assumptions on the sampling distribution of $Pr(H_1 \hspace{1pt}| \hspace{1pt}data)$. This broadly applicable method is economical and acknowledges the variability inherent to simulation-based design.

The remainder of this article is structured as follows. We introduce background information and notation in Section \ref{sec:intro}. In Section \ref{sec:multi}, we develop theory for a proxy to the sampling distribution of posterior probabilities, and we propose a method in Section \ref{sec:multi.alg} that adapts this theory to determine which $(n, \gamma)$ combination minimizes the sample size $n$ while satisfying criteria for power and the type I error rate. This procedure estimates the sampling distributions of posterior probabilities at only two sample sizes. Given bootstrap samples from these estimated sampling distributions, we construct confidence intervals for the optimal sample size and critical value. We illustrate the use of our design framework with examples based on semaglutide development in Section \ref{sec:multi.ex}. In Section \ref{sec:contour}, we repurpose the posterior probabilities used to find the optimal $(n, \gamma)$ combination to create contour plots that facilitate the investigation of various $n$ and $\gamma$ values. Section \ref{sec:disc} concludes with a summary and discussion of extensions to this work. 

  \section{Preliminaries}\label{sec:intro}


Our design framework generally represents data from a random, to-be-observed sample of $n$ potentially multivariate observations as $\boldsymbol{W}^{_{(n)}} = \{\boldsymbol{W}_{i} \}_{i = 1}^n$. The observed data are denoted by $\boldsymbol{w}^{_{(n)}}$. In regression settings, $\boldsymbol{W}_{i}$ may consist of a scalar response $Y_i$ and a vector of explanatory covariates $\boldsymbol{X}_i$. For two-group comparisons, $\boldsymbol{W}^{_{(n)}}$ consists of $n = n_A + n_B$ observations, where the first $n_A$ observations are from group A and the final $n_B$ observations are from group B. We consider fixed treatment allocation in these settings such that $n_A = \lfloor qn_B \rceil$ for some $q > 0$, but this constant $q$ is not incorporated into $\boldsymbol{W}^{_{(n)}}$. We assume that each observation in $\boldsymbol{W}^{_{(n)}}$ is generated independently according to the model $f(\boldsymbol{w}; \boldsymbol{\eta})$, where $\boldsymbol{\eta}$ denotes a vector of parameters. 

The estimand $\theta$ is typically specified as a function $g(\cdot)$ of these parameters: $\theta = g(\boldsymbol{\eta})$. For posterior analyses, we consider interval hypotheses of the form $H_1: \theta \in \boldsymbol{\delta} = (\delta_L, \delta_U)$, where $-\infty \le \delta_L < \delta_U \le \infty$. The complementary hypothesis is $H_0: \theta \notin \boldsymbol{\delta}$. We use general notation for the interval $\boldsymbol{\delta}$ to accommodate a broad suite of hypothesis tests based on superiority, noninferiority, and practical equivalence \citep{spiegelhalter1994bayesian, spiegelhalter2004bayesian}. We assume larger $\theta$ values are preferred to introduce several such hypotheses. In that case, the interval $(\delta_L, \delta_U) = (0, \infty)$ facilitates one-sided hypothesis tests based on superiority. The interval endpoint $\delta_L$ might instead take a negative (positive) value to assess hypotheses based on noninferiority (practical superiority). For an equivalence test, both interval endpoints of $\boldsymbol{\delta}$ would take finite values.


Algorithm \ref{alg.int} details a simulation-based procedure to estimate the sampling distributions of posterior probabilities at a given sample size $n$. We now discuss several design inputs for this algorithm. We characterize data generation for $\boldsymbol{W}^{_{(n)}}$ using the model $f^+(\boldsymbol{w}; \boldsymbol{\eta}^+)$. We have that $\boldsymbol{\eta} \subseteq \boldsymbol{\eta}^+$ since additional parameters beyond those needed to specify $\theta$ may be required to generate data. For instance, regression settings require additional parameters because those used to generate the explanatory covariates $\{\boldsymbol{X}_i\}_{i=1}^n$ are not part of the regression model. The model $f^+(\boldsymbol{w}; \boldsymbol{\eta}^+)$ generalizes $f(\boldsymbol{w}; \boldsymbol{\eta})$ to take these additional parameters. The prior $p(\boldsymbol{\eta})$ is used to induce the posterior of $\theta = g(\boldsymbol{\eta})$. We must also specify the number of simulation repetitions $m$ for Algorithm \ref{alg.int}, and we provide guidance for this choice when quantifying simulation variability in Section \ref{sec:multi.alg}.

The final design inputs that we must specify are $\Psi_0$ and $\Psi_1$, probability models that characterize how $\boldsymbol{\eta}^+$ values are drawn in each simulation repetition $r = 1, \dots, m$. The probability model $\Psi_j$ outputs $\boldsymbol{\eta}^+$ values that correspond to $H_j$, $j = 0$, 1. The inputs $\Psi_0$ and $\Psi_1$ could be viewed as \emph{design} priors \citep{de2007using,berry2010bayesian,gubbiotti2011bayesian} that differ from the \emph{analysis} prior $p(\boldsymbol{\eta})$. If $\Psi_j$ is degenerate, then the values $\{\boldsymbol{\eta}^+_{j, r}\}_{r = 1}^m$ generated across all $m$ simulation repetitions in Line 4 of Algorithm \ref{alg.int} are identical. If $\Psi_1$ is not degenerate and incorporates uncertainty about the parametric assumptions used to generate data, power is commonly referred to as \emph{assurance} \citep{o2001bayesian}.

\begin{algorithm}
\caption{Sampling Distribution Estimation}
\label{alg.int}

\begin{algorithmic}[1]
\setstretch{1}
\Procedure{Estimate}{$f^+(\cdot)$, $g(\cdot)$, $\boldsymbol{\delta}$, $p(\boldsymbol{\eta})$, $n$, $q$, $m$, $\Psi_0$, $\Psi_1$}
\For{$j$ in \{0,1\}}
\For{$r$ in 1:$m$}
    \State Generate $\boldsymbol{\eta}^+_{j, r} \sim \Psi_j$ and $\boldsymbol{w}^{_{(n)}}_{j, r} \sim f^+(\boldsymbol{\eta}^+_{j, r})$ 
    \State Compute estimate $\widehat{Pr}(H_1 \hspace{1pt}| \hspace{1pt}\boldsymbol{w}^{_{(n)}}_{j, r})$ 
    \EndFor
    \EndFor
    \State \Return $\{\widehat{Pr}(H_1 \hspace{1pt}| \hspace{1pt}\boldsymbol{w}^{_{(n)}}_{1, r})\}_{r = 1}^m$ and $\{\widehat{Pr}(H_1 \hspace{1pt}| \hspace{1pt}\boldsymbol{w}^{_{(n)}}_{0, r})\}_{r = 1}^m$
\EndProcedure

\end{algorithmic}
\end{algorithm}

Line 4 of Algorithm \ref{alg.int} also generates a sample $\boldsymbol{w}^{_{(n)}}_{j, r}$ in each simulation repetition given the value for $\boldsymbol{\eta}^+_{j, r}$ drawn from $\Psi_j$. This collection of estimates $\{\widehat{Pr}(H_1 \hspace{1pt}| \hspace{1pt}\boldsymbol{w}^{_{(n)}}_{j, r})\}_{r = 1}^m$ is used to estimate the sampling distribution of posterior probabilities under the hypothesis $H_j$. Power and the type I error rate can respectively be estimated as
\begin{equation}\label{eq:oc.est}
\dfrac{1}{m}\sum_{r=1}^m\mathbb{I}\left\{\widehat{Pr}(H_1 \hspace{1pt}| \hspace{1pt}\boldsymbol{w}^{_{(n)}}_{1, r}) \ge \gamma\right\} ~~~ \text{and} ~~~ \dfrac{1}{m}\sum_{r=1}^m\mathbb{I}\left\{\widehat{Pr}(H_1 \hspace{1pt}| \hspace{1pt}\boldsymbol{w}^{_{(n)}}_{0, r}) \ge \gamma\right\}.
\end{equation} 
To determine whether criteria for the operating characteristics are satisfied, we introduce the notation $\xi(a,b)$ to denote the $a^{\text{th}}$ order statistic of the collection of observations $b$. For an $(n, \gamma)$ combination, the estimated power in (\ref{eq:oc.est}) is at least $1 - \beta$ if and only if $\xi_1 = \xi(\lfloor  m \beta \rfloor, \{\widehat{Pr}(H_1 \hspace{1pt}| \hspace{1pt}\boldsymbol{w}^{_{(n)}}_{1, r})\}_{r = 1}^m) \ge \gamma$. The estimated type I error rate in (\ref{eq:oc.est}) is at most $\alpha$ if and only if $\xi_0 = \xi(\lceil  m(1 - \alpha )\rceil, \{\widehat{Pr}(H_1 \hspace{1pt}| \hspace{1pt}\boldsymbol{w}^{_{(n)}}_{0, r})\}_{r = 1}^m) \le \gamma$.

We now describe how $n$ and $\gamma$ impact the operating characteristics of a posterior analysis. The sampling distributions of posterior probabilities change with $n$. As $n \rightarrow \infty$, the sampling distribution of posterior probabilities under $H_1$ converges to a point mass at 1 under standard regularity conditions \citep{vaart1998bvm}. A sample size $n$ is sufficiently large if and only if $\xi_0 \le \xi_1$ since both criteria in (\ref{eq:oc.est}) are satisfied for any $\gamma \in [\xi_0, \xi_1]$. Otherwise, there is no value of $\gamma$ such that the estimated power is at least $1 - \beta$ and the estimated type I error rate is at most $\alpha$. For any $\Psi_0$ assigning all weight to $\boldsymbol{\eta}_{j,r} \subseteq \boldsymbol{\eta}^+_{j,r}$ values such that $\theta_{j,r} = g(\boldsymbol{\eta}_{j,r})$ is an endpoint of $\boldsymbol{\delta}$, the sampling distribution of posterior probabilities under $H_0$ converges to the standard uniform distribution as $n \rightarrow \infty$ under weak conditions \citep{bernardo2009bayesian}.  In such cases, choosing $\gamma = 1 - \alpha$ yields a type I error rate of approximately $\alpha$ for large sample sizes. However, this choice for $\gamma$ may not maintain the desired type I error rate with finite samples or certain nondegenerate probability models $\Psi_0$ as demonstrated in this paper. 

To better satisfy criteria for both operating characteristics, our design procedures make optimal choices for both $n$ and $\gamma$ instead of selecting a sample size $n$ to achieve a study power of $1 - \beta$ given a predetermined critical value of $\gamma = 1 - \alpha$. These choices for $n$ and $\gamma$ are informed by estimating the sampling distributions of posterior probabilities under $H_1$ ($H_0$) to estimate $\xi_1$ ($\xi_0$). The procedure in Algorithm \ref{alg.int} requires independent implementation for each sample size $n$ that we consider. This process is often computationally intensive, but we could reduce the computational burden by using the estimated sampling distributions under $H_1$ ($H_0$) for previously considered sample sizes to estimate $\xi_1$ ($\xi_0$) for new $n$ values. We could use this process to explore $(n, \gamma)$ combinations with substantially fewer simulation repetitions. We propose such a method for the design of posterior analyses in this paper and begin its development in Section \ref{sec:multi}. 

\section{A Proxy for the Sampling Distribution}\label{sec:multi}

 Design methods to control the operating characteristics of posterior analyses require that we estimate the sampling distribution of posterior probabilities for various sample sizes $n$. We approximate such sampling distributions by generating data $\boldsymbol{w}^{_{(n)}}$ using the straightforward process in Algorithm \ref{alg.int}. For theoretical development, we create a proxy for these sampling distributions. These proxies motivate -- but are not used in -- our design methods proposed in Section \ref{sec:multi.alg}. A detailed understanding of these proxies is not necessary to appreciate the practical benefits of our design methods that are simple and straightforward to implement, but the proxies are needed for the theory that underpins the proposed methodology. 
 
 Our proxies make use of the regularity conditions listed in Appendix A of the supplement. Appendix A.1 details the four necessary assumptions to invoke the Bernstein-von Mises (BvM) theorem \citep{vaart1998bvm}. The first three assumptions are weaker than the regularity conditions for the asymptotic normality of the maximum likelihood estimator (MLE) \citep{lehmann1998theory}, which are listed in Appendix A.2. By the BvM theorem, a large-sample approximation to the posterior of $\theta~| ~\boldsymbol{w}^{_{(n)}}_{j, r}$ is $\mathcal{N}(\hat{\theta}^{_{(n)}}_{j,r}, \mathcal{I}(\theta_{j,r})^{-1}/n)$ \citep{vaart1998bvm}. Here, $\hat{\theta}^{_{(n)}}_{j,r}$ is the maximum likelihood estimate of $\theta$, $\mathcal{I}(\cdot)$ is the Fisher information, and $\theta_{j,r} = g(\boldsymbol{\eta}_{j,r})$ corresponds to $\boldsymbol{\eta}_{j,r} \subseteq \boldsymbol{\eta}^+_{j,r} \sim \Psi_j$. The approximate sampling distribution of $\hat{\theta}^{_{(n)}}~|~\boldsymbol{\eta}^+ = \boldsymbol{\eta}^+_{j,r}$ is $\mathcal{N}(\theta_{j,r}, \mathcal{I}(\theta_{j,r})^{-1}/n)$. We can simulate one realization from this approximate sampling distribution using cumulative distribution function (CDF) inversion with a point $u_{j,r} \in [0,1]$:  
   \begin{equation}\label{eq:cdf.inv}
\hat{\theta}^{_{(n)}}_{j,r} = \theta_{j,r} + \Phi^{-1}(u_{j,r})\sqrt{\frac{\mathcal{I}(\theta_{j,r})^{-1}}{n}},
\end{equation} 
where $\Phi(\cdot)$ is the standard normal CDF. Thus, we require a pseudorandom sequence of $m$ points $\{u_{j,r}\}_{r = 1}^m \in [0,1]$ to simulate from the approximate distribution of $\hat{\theta}^{_{(n)}}$ under $H_j$. This sample allows us to estimate a proxy for the sampling distribution of the relevant posterior probability. We let 
      \begin{equation}\label{eq:proxy}
p^{_{(n)}}_{\delta, j, r} = 
   \Phi\left(\dfrac{\delta - \hat{\theta}^{_{(n)}}_{j,r}}{\sqrt{\mathcal{I}(\theta_{j,r})^{-1}/n}}\right) = \Phi\left(\dfrac{\delta - \theta_{j,r}}{\sqrt{\mathcal{I}(\theta_{j,r})^{-1}}}\sqrt{n} - \Phi^{-1}(u_{j,r})\right)
\end{equation} 
be a large-sample approximation to $Pr(\theta < \delta\hspace{1pt}| \hspace{1pt}\boldsymbol{w}^{_{(n)}}_{j, r})$ obtained with the point $u_{j,r} \in [0,1]$ given $\boldsymbol{\eta}^+ = \boldsymbol{\eta}^+_{j,r}$. This approximation is based on the BvM theorem and (\ref{eq:cdf.inv}). The rightmost equality in (\ref{eq:proxy}) underscores how $p^{_{(n)}}_{\delta, j, r}$ changes with the sample size $n$. We emphasize that the probability $p^{_{(n)}}_{\delta, j, r}$ depends on the point $u_{j,r}$ and parameter value $\boldsymbol{\eta}^+_{j,r}$ through the subscripts $j$ and $r$. 

For theoretical purposes, a proxy sampling distribution of posterior probabilities under $H_j$ could be constructed using a collection of $p^{_{(n)}}_{\boldsymbol{\delta}, j, r} = p^{_{(n)}}_{\delta_U, j, r} - p^{_{(n)}}_{\delta_L, j, r}$ values corresponding to $\{u_{j,r}\}_{r = 1}^m \in [0,1]$ and $\{\boldsymbol{\eta}_{j,r}^+\}_{r=1}^m \sim \Psi_j$. Theorem 1 of \citet{hagar2023fast} proved that under the conditions in Appendix A, the total variation distance between the sampling distribution of $p^{_{(n)}}_{\boldsymbol{\delta}, j, r}$ and that of ${Pr}(H_1 \hspace{1pt}| \hspace{1pt}\boldsymbol{w}^{_{(n)}}_{j, r})$ converges in probability to 0 as $n \rightarrow \infty$. The estimates of power and the type I error rate based on the (proxy) sampling distribution of $p^{_{(n)}}_{\boldsymbol{\delta}, j, r}$ are therefore consistent as $n \rightarrow \infty$. This consistency is adequate for theoretical consideration of the proxy sampling distribution, but we do not use it in Section \ref{sec:multi.alg}. The corresponding estimates based on the (true) sampling distribution of ${Pr}(H_1 \hspace{1pt}| \hspace{1pt}\boldsymbol{w}^{_{(n)}}_{j, r})$ obtained via Algorithm \ref{alg.int} are instead unbiased for finite $n$ so long as the posterior approximation method does not introduce bias. 

Theorem \ref{thm1} below is original to this paper. It provides guidance concerning how to economically assess the operating characteristics of a posterior analysis at a broad range of sample sizes. In particular, Theorem \ref{thm1} guarantees that the logit of $p^{_{(n)}}_{\boldsymbol{\delta}, j, r}$ constructed using posterior probabilities in (\ref{eq:proxy}) is an approximately linear function of $n$; hence, exploration of the $(n, \gamma)$-space can be achieved by estimating the sampling distributions of posterior probabilities under $H_0$ and $H_1$ at \emph{only two} values of $n$. The probabilities $p^{_{(n)}}_{\boldsymbol{\delta}, j, r}$ in (\ref{eq:proxy}) depend on the model $f^+(\boldsymbol{w}; \boldsymbol{\eta}^+_{j,r})$, the sample size $n$, and the point $u_{j,r}$. We fix the point $u_{j,r}$ and let the sample size $n$ vary. When the point $u_{j,r}$ and model $f^+(\boldsymbol{w}; \boldsymbol{\eta}^+_{j,r})$ are fixed, $p^{_{(n)}}_{\boldsymbol{\delta}, j, r}$ is a deterministic function of $n$. 
\begin{theorem}\label{thm1}
    For any $\boldsymbol{\eta}^+_{j,r} \sim \Psi_j$, let the model $f(\boldsymbol{w};\boldsymbol{\eta}^+_{j,r})$ satisfy the conditions in Appendix A.2 and the prior $p(\boldsymbol{\eta})$ satisfy the conditions in Appendix A.1.
    Define $\emph{logit}(x) = \emph{log}(x) - \emph{log}(1-x)$
    and $a(\delta, \theta_{j,r}) = (\delta - \theta_{j,r})/\sqrt{\mathcal{I}(\theta_{j,r})^{-1}}$. For a given point $u_{j,r} \in [0,1]$, the function $p^{_{(n)}}_{\delta, j, r}$ in (\ref{eq:proxy}) is such that
$$\lim\limits_{n \rightarrow \infty} \dfrac{d}{dn}~\emph{logit}\left[p^{_{(n)}}_{\delta_U, j, r} - p^{_{(n)}}_{\delta_L, j, r}\right]= (0.5 - \mathbb{I}\{\theta_{j,r} \notin (\delta_L, \delta_U)\})\times\emph{min}\{a(\delta_U, \theta_{j,r})^2, a(\delta_L, \theta_{j,r})^2\}.$$ 
\end{theorem} 
Theorem \ref{thm1} is proved in Appendix B of the supplement, but we consider its practical implications here. For the proxy sampling distributions, the linear approximation to $l^{_{(n)}}_{\boldsymbol{\delta}, j, r} = \text{logit}(p^{_{(n)}}_{\boldsymbol{\delta}, j, r})$ as a function of $n$ is a good global approximation for large sample sizes. This linear approximation should also be locally suitable for a range of smaller sample sizes.
Therefore, the quantiles of the sampling distribution of $l^{_{(n)}}_{\boldsymbol{\delta}, j, r}$ change linearly as a function of $n$ when $\Psi_j$ is degenerate. We exploit this linear trend in the proxy sampling distributions to flexibly model logits of posterior probabilities as linear functions of $n$ when independently simulating samples $\boldsymbol{w}^{_{(n)}}$ as in Algorithm \ref{alg.int} -- even when $\Psi_j$ is nondegenerate. Since we only use the limiting slopes from Theorem \ref{thm1} to initialize our method, we do not require very large sample sizes to apply our methodology with the true sampling distributions in Section \ref{sec:multi.alg}. This will be illustrated in Section \ref{sec:multi.ex}.


\section{Economical Assessment of Operating Characteristics}\label{sec:multi.alg}


The methods we propose to design posterior analyses can be broadly applied when the conditions in Appendix A are satisfied. We generalize the results from Theorem \ref{thm1} to develop methodology in Algorithm \ref{alg4} that is easily implemented and performs well with moderate to large sample sizes. Algorithm \ref{alg4} allows users to efficiently explore the sample size space to find the $(n, \gamma)$ combination that minimizes the sample size while satisfying both criteria in (\ref{eq:oc.est}). Our approach involves estimating the sampling distributions of posterior probabilities at only two sample sizes: $n_0$ and $n_1$.

\begin{algorithm}
\caption{Procedure to Determine Optimal Sample Size and Critical Value}
\label{alg4}

\begin{algorithmic}[1]
\setstretch{1}
\Procedure{Optimize}{$f^+(\cdot)$, $g(\cdot)$, $\boldsymbol{\delta}$, $p(\boldsymbol{\eta})$, $q$, $\alpha$, $\beta$, $m$, $\Psi_0$, $\Psi_1$}
\State Select $n_0$ to achieve power $1-\beta$ for $\gamma = 1- \alpha$ based  on the BvM theorem
\For{$j$ in \{0, $1$\}}
\State Estimate $\{\widehat{Pr}(H_1 \hspace{1pt}| \hspace{1pt}\boldsymbol{w}^{_{(n_0)}}_{j, r})\}_{r = 1}^m$ via Algorithm \ref{alg.int} and their logits $\{\hat{l}^{_{(n_0)}}_{\boldsymbol{\delta}, j, r}\}_{r = 1}^m$
\For{$r$ in $1$:$m$}
  \State Use the line $\hat{L}^{_{(n)}}_{\boldsymbol{\delta}, j, r}$ passing through $(n_0, \hat{l}^{_{(n_0)}}_{\boldsymbol{\delta}, j, r})$ with the slope from  Theorem \ref{thm1} to get  $\hat{l}^{_{(n)}}_{\boldsymbol{\delta}, j, r}$ for \linebreak \hspace*{42pt} other sample sizes
 \EndFor
 \EndFor
\State Use binary search to find $n_1$, the smallest $n$ such that  $\xi$($\lfloor  m \beta \rfloor$, $\{\hat{l}^{_{(n)}}_{\boldsymbol{\delta}, 1,r}\}_{r=1}^{m}$) $\ge$ $\xi$($\lceil  m(1 - \alpha) \rceil$, $\{\hat{l}^{_{(n)}}_{\boldsymbol{\delta}, 0,r}\}_{r = 1}^{m}$)
 \For{$j$ in \{0, $1$\}}
\State Estimate $\{\widehat{Pr}(H_1 \hspace{1pt}| \hspace{1pt}\boldsymbol{w}^{_{(n_1)}}_{j, r})\}_{r = 1}^m$ via Algorithm \ref{alg.int} and their logits $\{\hat{l}^{_{(n_1)}}_{\boldsymbol{\delta}, j, r}\}_{r = 1}^m$
\For{$r$ in $1$:$m$}
  \State Use the line $\hat{L}^{_{(n)}}_{\boldsymbol{\delta}, j, r}$ passing through $(n_0, \xi(r, \{\hat{l}^{_{(n_0)}}_{\boldsymbol{\delta}, j, r}\}_{r=1}^{m}))$ and  $(n_1, \xi(r, \{\hat{l}^{_{(n_1)}}_{\boldsymbol{\delta}, j, r}\}_{r=1}^{m}))$ to get $\hat{l}^{_{(n)}}_{\boldsymbol{\delta}, j, r}$ \linebreak \hspace*{42pt} for other sample sizes
 \EndFor
 \EndFor
\State Use binary search to find $n_2$, the smallest $n$ such that  $\xi$($\lfloor  m \beta \rfloor$, $\{\hat{l}^{_{(n)}}_{\boldsymbol{\delta}, 1,r}\}_{r=1}^{m}$) $\ge$ $\xi$($\lceil  m(1 - \alpha) \rceil$, $\{\hat{l}^{_{(n)}}_{\boldsymbol{\delta}, 0,r}\}_{r = 1}^{m}$)
 \State \Return $n_2$ as recommended $n$ and $\xi$($\lceil  m(1 - \alpha) \rceil$, $\{\hat{p}^{_{(n_2)}}_{\boldsymbol{\delta}, 0, r}\}_{r = 1}^{m}$) as $\gamma$

\EndProcedure

\end{algorithmic}
\end{algorithm}

We elaborate on several steps in Algorithm \ref{alg4} below. We choose the initial sample size $n_0$ in Line 2 under the assumption that the posterior of $\theta$ is $\mathcal{N}(\hat{\theta}^{_{(n)}}, \mathcal{I}(\theta_*)^{-1}/n)$, where $\hat{\theta}^{_{(n)}} \sim \mathcal{N}({\theta}_*, \mathcal{I}(\theta_*)^{-1}/n)$ is the MLE and $\theta_*$ is the median value of $\theta$ induced by $\boldsymbol{\eta}^+ \sim \Psi_1$. This sample size provides a suitable starting point, but it may differ materially from the optimal $n$ if $\Psi_1$ is nondegenerate or the large-sample approximations based on the theory in Appendix A are not suitable. One strength of our methodology is its flexibility: Algorithm \ref{alg4} can readily
be integrated with any computational or analytical method used to estimate the posterior probabilities in Line 4. If a computational method is used to generate posterior samples, we recommend calculating posterior probabilities using a nonparametric kernel density estimate of the posterior so that the logits of all probabilities are finite. The notation $\hat{l}^{_{(n)}}_{\boldsymbol{\delta}, j, r}$ is also introduced in Line 4. These logits and their posterior probabilities $\hat{p}^{_{(n)}}_{\boldsymbol{\delta}, j, r}$ from the true sampling distribution leverage independently generated samples $\boldsymbol{w}^{_{(n)}}_{j, r}$ for each hypothesis $j$ and simulation repetition $r$. Unlike for $l^{_{(n)}}_{\boldsymbol{\delta}, j, r}$ from the proxy sampling distribution in Theorem \ref{thm1}, there is no relationship between the $\hat{l}^{_{(n)}}_{\boldsymbol{\delta}, j, r}$ values corresponding to two different sample sizes that happen to have the same indices for $j$ and $r$.


To choose a sample size $n_1$ that improves on the initial one, we construct linear approximations to the logits of posterior probabilities as a function of $n$ using the limiting slopes from Theorem \ref{thm1} in Line 6. For moderate $n$, the limiting slopes for $l^{_{(n)}}_{\boldsymbol{\delta}, j, r}$ may not be accurate since Theorem \ref{thm1} relies on large-sample results -- including the approximate normality of the posterior that is asymptotically guaranteed when the conditions for the BvM theorem are satisfied. Thus, we only use those slopes in this initial phase of our method. If using analytical posterior approximation, considering the posterior of a monotonic transformation of $\theta$ may improve its normal approximation and the accuracy of the limiting slopes for moderate $n$. If computational posterior approximation is used, monotonic transformations of $\theta$ need not be considered. The order statistics in Line 7 are quickly calculated based on the linear approximations -- not by approximating posterior probabilities based on observed data. 

In Line 11 of Algorithm \ref{alg4}, we construct linear approximations to logits of posterior probabilities that are less reliant on large-sample results. These approximations use independent estimates of each sampling distribution (under $H_0$ and $H_1$) at the sample sizes $n_0$ and $n_1$, and they exploit the linear trend in the proxy sampling distribution quantiles discussed in Section \ref{sec:multi}. For the true sampling distributions, we sort the logits of the posterior probabilities estimated under $H_j$ at $n_0$ and $n_1$ and construct linear approximations using the same order statistic at both sample sizes. This approach is suitable when the true value of the estimand $\theta_{j,r}$ is similar for all $\boldsymbol{\eta}^+_{j,r} \sim \Psi_j$. When $\Psi_j$ is nondegenerate, the process in Line 11 can be modified: we instead split the logits of the posterior probabilities for each sample size into subgroups based on the order statistics of their $\theta_{j,r}$ values before constructing the linear approximations. 

In Line 12, we repeat the process in Line 7 with our improved linear approximations to obtain the final sample size recommendation $n_2$. These linear approximations yield unbiased estimates of the operating characteristics at $n_0$ and $n_1$. Thus, the suitability of the sample size recommendation $n_2$ only relies on the accuracy of the \emph{empirically} estimated slopes. The optimal critical value is the $\lceil  m(1 - \alpha) \rceil^{\text{th}}$ order statistic of $\{\hat{l}^{_{(n_2)}}_{\boldsymbol{\delta}, 0, r}\}_{r=1}^{m}$ on the probability scale. If $n_1$ and $n_2$ differ greatly, the large-sample approximations used in Lines 2 and 6 may not be suitable. We could approximate the sampling distributions of posterior probabilities at $n_2$ using Algorithm \ref{alg.int} in that event. Lines 8 to 12 of Algorithm \ref{alg4} could be rerun using $n_1$ and $n_2$ instead of $n_0$ and $n_1$, respectively. However, we have found that it is generally not necessary to approximate the sampling distributions of posterior probabilities at a third sample size. 

We now describe how to construct bootstrap confidence intervals for the optimal $n$ and $\gamma$ values given a single implementation of Algorithm \ref{alg4}.  In Algorithm \ref{alg4}, we obtain four estimates of the sampling distribution of posterior probabilities: $\{\widehat{Pr}(H_1 \hspace{1pt}| \hspace{1pt}\boldsymbol{w}^{_{(n_0)}}_{j, r})\}_{r = 1}^m$ and $\{\widehat{Pr}(H_1 \hspace{1pt}| \hspace{1pt}\boldsymbol{w}^{_{(n_1)}}_{j, r})\}_{r = 1}^m$ for $j = 0, 1$. We independently obtain a bootstrap sample from each estimated sampling distribution by resampling with replacement. We use these four bootstrap sampling distribution estimates to obtain linear approximations to logits of posterior probabilities as a function of $n$ as in Line 11. These linear approximations obtained using the bootstrap samples give rise to a new $(n, \gamma)$ recommendation following the process in Lines 12 and 13. We repeat this procedure $M$ times to construct bootstrap confidence intervals using the percentile method \citep{efron1982jackknife}. 

These confidence intervals could be used to help select the number of simulation repetitions $m$. The U.S. Food and Drug Administration (FDA) recommends using at least $10^4$ simulation repetitions to estimate sampling distributions \citep{fda2019adaptive}. So, Algorithm \ref{alg4} could be run with an initial value of $m = 10^4$; if the resulting bootstrap confidence intervals for $n$ and $\gamma$ were not sufficiently precise, one could use Algorithm \ref{alg.int} with sample sizes $n_0$ and $n_1$ to augment the sampling distributions estimated in Lines 4 and 9 of Algorithm \ref{alg4}. This process to incrementally increase $m$ could be alternated with the bootstrap procedure until the confidence intervals for $n$ and $\gamma$ were narrow enough. Furthermore, this process is incredibly economical in that previously estimated posterior probabilities can be efficiently repurposed. We investigate the performance of Algorithm \ref{alg4} and the procedure to construct bootstrap confidence intervals when considering study design for several examples in Section \ref{sec:multi.ex}.

\section{Numerical Studies}\label{sec:multi.ex}

 \subsection{Example 1}\label{sec:multi.ex2}

 Here, we consider a two-group comparison based on a recent clinical trial for semaglutide development \citep{wilding2021once}. In this clinical trial, patients in groups A and B were respectively given a weekly semaglutide injection or placebo for 68 weeks.  A total of $n_A = 1306$ and $n_B = 655$ patients were enrolled in this study. While the primary outcome for this study concerned weight loss, the proportion of participants that experienced a serious adverse event (SAE) was also of interest. As detailed in \citet{wilding2021once}, 9.8\% and 6.4\% of patients respectively receiving the semaglutide and placebo experienced SAEs. We now suppose that we want to design a two-group comparison for an early clinical trial of a similar semaglutide medication. As part of this trial, we consider a Bayesian logistic regression model for the probability of experiencing an SAE. This model is such that $y_i \sim \text{BIN}(1, \pi_i)$, where $\text{logit}(\pi_i) = \beta_0 + \beta_1x_{1i} + \beta_2x_{2i}$, $y \in \{0, 1\}$ denotes whether an SAE is experienced, $x_1 = \mathbb{I}(\text{Group} = A)$ is the binary treatment indicator, $x_2$ is the patient's baseline weight in kilograms. We use this model for pedagogical purposes, but our methodology accommodates much more complex models so long as the conditions for Theorem \ref{thm1} are satisfied. 

 For this comparison, the characteristic of interest is $\theta = \text{exp}(\beta_1)$, the odds ratio (OR) of experiencing an SAE when taking the semaglutide compared to the placebo. The model parameters are $\boldsymbol{\eta} = \boldsymbol{\beta} = (\beta_0, \beta_1, \beta_2)$. We aim to support the hypothesis $H_1: \theta \in \boldsymbol{\delta} = (-\infty, 2)$, which would suggest the semaglutide does not increase the OR of experiencing an SAE enough to preclude further study of the semaglutide in later trials. We use a treatment allocation constant of $q = 2$ as in \citet{wilding2021once}. For this example, we specify $\Psi_0$ as a degenerate process such that $\boldsymbol{\beta}^+_{0,r} = (-2.71, \text{log}(2), 0.25)$ and $\{x_{2i}\}_{i = 1}^n \overset{\text{i.i.d.}}{\sim} \mathcal{N}(0, 1)$ after centering and scaling the baseline weight. The process $\Psi_1$ is the same as $\Psi_0$ except that $\boldsymbol{\beta}^+_{1,r} = (-2.71, \text{log}(1.25), 0.25)$. For this example, we independently join a $\mathcal{N}(-2.71, 1)$ prior for $\beta_0$ with $\mathcal{N}(0, 10^2)$ priors for $\beta_1$ and $\beta_2$. This marginal prior for $\beta_0$ is rather informative for illustration, but we have substantial information about the prevalence of SAEs for patients taking the placebo from past studies. The motivation for the choices in this paragraph is further discussed in Appendix C.1 of the supplement.
 
 We define criteria for the operating characteristics using $\alpha = 0.4$ and $\beta = 0.25$. While the criterion for the type I error rate is quite lenient, it is not uncommon to use larger values for $\alpha$ with secondary safety outcomes in early phases of clinical trials. Moreover, the values for $\alpha$, $\beta$, and the $\beta_1$ component of $\boldsymbol{\beta}^+_{1,r}$ were selected to ensure this example illustrates the performance of our method with smaller sample sizes. We also used $m = 10^5$. This value is 10 times larger than the FDA's minimum recommendation ($10^4$), and we use a large value of $m$ to contextualize these recommendations. This example is interesting because informative priors and smaller sample sizes are considered, so analytical sample size calculations based on limiting results may perform poorly.

  When using Algorithm \ref{alg4}, the optimal design was characterized by $(n_B, \gamma) = (73, 0.5651)$. At this $(n_B, \gamma)$ combination, power and the type I error rate were estimated as 0.7535 and 0.4016 using confirmatory simulations. We have that $n = n_A + n_B = 3n_B$ in this case. This optimal design took roughly 2.5 hours to obtain using parallelization with 72 cores. This process would have taken about 15 minutes if $m = 10^4$ were used. With $m = 10^5$, Algorithm \ref{alg4} required us to approximate $4\times10^5$ posterior probabilities. We approximated each posterior of $\theta$ using Markov chain Monte Carlo methods with $10^3$ posterior draws and 500 burnin iterations. This procedure could have been expedited if we used analytical posterior approximation or approximate Bayesian computation. Despite the long runtime for this example, Algorithm \ref{alg4} is 4 times faster than exploring the sample size space via binary search. The discrepancy in runtime between estimating the sampling distributions of posterior probabilities at only two samples sizes and thoroughly exploring the sample size space scales logarithmically as the recommended value of $n$ increases.

  With $M = 10^4$, we implemented the bootstrap procedure from Section \ref{sec:multi.alg} with each of the following sample sizes for the bootstrap resamples: $m^* = 10^4 \times \{1, 2.5, 5, 7.5, 10\}$. While $m^*$ and $m$ are typically equal, our numerical studies consider settings where $m^* < m$ to explore the precision of the bootstrap confidence intervals. Table \ref{tab:boot} details the 95\% bootstrap confidence intervals for $n_B$ and $\gamma$ obtained from this numerical study. First, we note that the confidence interval for $n_B$ is rather wide when $m^* = 10^4$. This interval includes total sample size recommendations for $n = 3n_B$ ranging between 195 and 237. Therefore, using only $m = 10^4$ simulation repetitions to estimate the sampling distributions may not always meaningfully inform sample size determination. The variability in this bootstrap confidence interval is not specific to our method in Algorithm \ref{alg4} -- it reflects the variability inherent to simulation-based design. This variability is often underreported because it is difficult to construct interval estimates for recommended sample sizes without repeatedly implementing simulation-based design methods. The theory in this paper allows us to explore the sample size space via numerical studies at only two sample sizes; the results in Table \ref{tab:boot} underscore the advantages of obtaining fewer high-quality estimates of the sampling distributions instead of many estimates of lower quality. Second, we note that none of the 95\% confidence intervals for $\gamma$ in Table \ref{tab:boot} include $1 - \alpha = 0.6$. 

    \begin{table}[t]
    \centering
\caption{95\% bootstrap confidence intervals for $n_B$ and $\gamma$ obtained with $M = 10^4$ and various values of $m^*$}
\label{tab:boot}
\begin{tabular}{@{}lccccc@{}}
\hline
& \multicolumn{5}{c}{$m^*$} \\[1pt]
\cline{2-6} \\[-6.1pt]
 & \multicolumn{1}{c}{$10^4$}
& \multicolumn{1}{c}{$2.5\times10^4$} & \multicolumn{1}{c}{$5\times10^4$} & \multicolumn{1}{c}{$7.5\times10^4$} & \multicolumn{1}{c}{$1\times10^5$}  \\
\hline
$n_B$ &  (65, 79)  & (69, 77) & (70, 76) & (71, 76) & (71, 75)  \\
        $\gamma$ &  (0.5578, 0.5715) & (0.5608, 0.5692) & (0.5620, 0.5681) & (0.5626, 0.5676) & (0.5629, 0.5672) \\ \hline
\end{tabular}
\end{table}

 An $n_0$ value of 93 for group B was obtained in Line 2 of Algorithm \ref{alg4} for this example. For the $(n_B, \gamma)$ combination of (93, 0.6), the estimated type I error rate and power for this design were respectively 0.3622 and 0.7459. The type I error rate estimated via simulation is substantially less than $\alpha = 0.4$, but the power criterion is not satisfied. When taking $n_B = 97$ recommended by a modified version of Algorithm \ref{alg4} with fixed $\gamma = 0.6$, we obtained estimates for power and the type I error rate of $0.7482$ and $0.3611$. There is a considerable discrepancy between these sample sizes and the value for $n_B$ of 73 recommended by Algorithm \ref{alg4}. This discrepancy is driven in part by the informative prior that was used for $\beta_0$. Furthermore, the expected number of SAEs in the placebo group was less than 5 when $n_B = 73$, so large-sample normal approximations to the binomial distribution used in analytical sample size calculations may be inaccurate. Nevertheless, Algorithm \ref{alg4} can readily be used to select optimal $(n, \gamma)$ combinations that satisfy both criteria in (\ref{eq:oc.est}), so it is a valuable alternative to naive calculations based on large-sample results and fixed $\gamma$ values.


\subsection{Example 2}\label{sec:multi.ex1}


We now reconsider the clinical trial detailed in \citet{wilding2021once}. One primary outcome in that trial was the percentage change in body weight over the course of the study. The patients who were given the semaglutide lost an average of 12.4\% more of their initial weight than the patients who were given the placebo. We again suppose that we want to design a two-group comparison for an early clinical trial of a similar semaglutide medication. The regression model that we consider takes the following form: $y_i = \beta_0 + \beta_1x_{1i} + \beta_2x_{2i} + \varepsilon_i$, where $y$ is percentage change in body weight, $x_1 = \mathbb{I}(\text{Group} = A)$ is the binary treatment indicator, $x_2$ is the patient's baseline waist circumference in centimeters, and $\varepsilon_i \sim \mathcal{N}(0, \sigma^2_{\varepsilon})$ are independent error terms. 

For this comparison, the characteristic of interest is $\theta = \beta_1$, the increased amount of weight loss (in \%) associated with taking the semiglutide injections. The model parameters are $\boldsymbol{\eta} = (\boldsymbol{\beta}, \sigma^2_{\varepsilon})$, where $\boldsymbol{\beta} = (\beta_0, \beta_1, \beta_2)$. We aim to support the hypothesis $H_1: \theta \in \boldsymbol{\delta} = (5, \infty)$, which would suggest the semaglutide yields substantial weight loss of at least 5\% more than the placebo to offset treatment side effects. We again use a treatment allocation constant of $q = 2$. For this example, we specify $\Psi_0$ as a degenerate process such that $\boldsymbol{\beta}^+_{0,r} = (-25.75, 5, 0.25)$, $\{x_{2i}\}_{i = 1}^n \overset{\text{i.i.d.}}{\sim} \mathcal{N}(115, 14.5^2)$, and $\{\varepsilon_i\}_{i = 1}^n \overset{\text{i.i.d.}}{\sim} \mathcal{N}(0, 10.07^2)$. These choices are justified using summary statistics from \citet{wilding2021once} in Appendix C.1 of the supplement. The process $\Psi_1$ is the same as $\Psi_0$ except the $\beta_1$ component of $\boldsymbol{\beta}^+_{1,r}  \overset{\text{i.i.d.}}{\sim} \mathcal{U}(9, 12)$. For this example, we use an uninformative conjugate normal-inverse-gamma prior $p(\boldsymbol{\eta})$ with the following parameters: $\boldsymbol{\mu}_0 = (0,0,0)$, $\boldsymbol{\lambda}_0 = 0.01 \times \mathbb{I}_{3}$, $a_0 = 1$, and $b_0 = 1$ such that $\mathbb{I}_{3}$ is the $3 \times 3$ identity matrix. We also use $\alpha = 0.05$, $\beta = 0.2$, and $m = 10^4$ for illustration.

Since $\Psi_1$ is not degenerate, we split the logits of the posterior probabilities under $H_1$ for each sample size into 10 subgroups based on the order statistics of their $\theta_{j,r}$ values before sorting them. We do not need to split the posterior probabilities under $H_0$ since $\Psi_0$ is degenerate. This example is interesting because our sampling distribution of posterior probabilities under $H_1$ is an infinite mixture of sampling distributions conditional on the $\beta_1$ component of $\boldsymbol{\beta}^+_{1,r}$. This result regularly holds true when $\Psi_1$ is nondegenerate and we consider assurance. While we only considered degenerate $\Psi_0$ processes in this paper, our framework does accommodate uncertainty in the data generation process under $H_0$. Once again, we have specified this example to consider a setting with small sample sizes where asymptotic approximations may perform poorly. The use of conjugate priors with this example also allows us to implement Algorithm \ref{alg4} many times to explore the coverage properties of the bootstrap confidence intervals for $n_B$ and $\gamma$. 

 When using Algorithm \ref{alg4}, the optimal design was characterized by $(n_B, \gamma) = (35, 0.9561)$. Again, we have that $n = 3n_B$. This optimal design took less than 4 seconds on a standard laptop without parallelization to obtain. For this example, it would take 3 times as long to explore the sample size space using binary search. This discrepancy would be much more pronounced for larger sample size recommendations. We repeated the process of determining the optimal design for this example 1000 times, which gave rise to 95\% confidence intervals for $n_B$ and $\gamma$ of (34, 36) and (0.9535, 0.9595). It is not computationally feasible to obtain these confidence intervals based on repeated implementation of Algorithm \ref{alg4} for more complex models, which is why we generally recommend using the bootstrap procedure in Section \ref{sec:multi.alg}. We note that this confidence interval for $\gamma$ excludes $1 - \alpha = 0.95$. The median recommendations for $n_B$ and $\gamma$ across these 1000 sample size calculations were 35 and 0.9564, respectively. At this $(n_B, \gamma)$ combination, power and the type I error rate were estimated as 0.8029 and 0.0500 using intensive simulation, which verified that $(35, 0.9564)$ is the true optimal $(n_B, \gamma)$ combination.

 We used the 1000 repetitions of Algorithm \ref{alg4} to evaluate the coverage properties of the bootstrap confidence intervals for $n_B$ and $\gamma$ that are feasible to create in practice. We implemented the bootstrap procedure from Section \ref{sec:multi.alg} alongside each repetition of Algorithm \ref{alg4} for this example with $m^* = m = 10^4$ and $M = 10^3$. This process gave rise to 1000 95\% bootstrap confidence intervals for $n_B$ and $\gamma$. 99.6\% of these confidence intervals for $n_B$ contained 35, and 96.1\% of these confidence intervals for $\gamma$ contained 0.9564. The discrete nature of the sample size caused these intervals to have coverage that exceeds the nominal level of 0.95. To illustrate this phenomenon, we note that the linear approximations from Algorithm \ref{alg4} allow us to approximate sampling distributions of posterior probabilities at noninteger values of $n_B$. 
 
 We then implemented a modified version of Algorithm \ref{alg4} where $n_2$ in Line 12 is found to the nearest hundredth. We made the same modification to the bootstrap procedure in Section \ref{sec:multi.alg} to allow noninteger $n_B$ recommendations. We used this procedure to obtain 1000 noninteger $(n_B, \gamma)$ recommendations, each of which was accompanied by a bootstrap confidence interval. The median recommendations for $n_B$ and $\gamma$ across these 1000 sample size calculations were 34.73 and 0.9565, respectively. 94.9\% of these new confidence intervals for $n_B$ contained 34.73, and 95.0\% of these new confidence intervals for $\gamma$ contained 0.9565. In practice, we take the $(n_B, \gamma)$ combination corresponding to the ceiling of those $n_B$ recommendations, so any bootstrap confidence interval with a lower endpoint in $(34.73, 35]$ or an upper endpoint in $(34, 34.73)$ would count as covering the true optimal value for $n_B$ after rounding up. For this reason, the estimated coverage for the confidence intervals was closer to the nominal value of 0.95 when allowing for noninteger sample sizes. We still, however, recommend constructing the bootstrap confidence intervals with integer sample sizes since rounding up cannot lead to systematically anti-conservative confidence intervals. The degree of conservatism of those bootstrap confidence intervals is more pronounced when $m$ is large enough to ensure the confidence interval for $n_B$ contains only several sample sizes.   
 


 
 An $n_0$ value of 32 for group B informed by a $\theta_*$ value of 10.5 was obtained in Line 2 of Algorithm \ref{alg4} for this example. For $\theta_*$ values of 9 and 12 that correspond to the extremes of the $\mathcal{U}(9, 12)$ distribution, the recommended $n_B$ value was respectively 59 and 20. It is difficult to choose $n_B$ analytically when $\Psi_1$ is nondegenerate because it is difficult to account for the sampling distribution of posterior probabilities under $H_1$ being a mixture distribution. For the $(n_B, \gamma)$ combination of (32, 0.95), the estimated type I error rate and power for this design were respectively 0.0573 and 0.7916. Thus, neither the criterion for power nor the type I error rate are satisfied. When taking $n_B = 33$ recommended by a modified version of Algorithm \ref{alg4} with fixed $\gamma = 0.95$, we obtained estimates for power and the type I error rate of $0.8012$ and $0.0571$. In this case, the power criterion is satisfied but the type I error criterion is not. These discrepancies produced by naive alternatives further emphasize the value of the proposed methodology.
 

\section{Contour Plots for Design Criteria Exploration}\label{sec:contour}

While Algorithm \ref{alg4} returns the $(n, \gamma)$ combination that minimizes the sample size $n$ while satisfying both criteria in (\ref{eq:oc.est}), practitioners may want to explore multiple designs that are similar to the optimal one. We obtained linear approximations $\hat{L}^{_{(n)}}_{\boldsymbol{\delta}, j, r}$ in Line 11 of Algorithm \ref{alg4} using estimates of the sampling distributions of posterior probabilities under $H_0$ and $H_1$ at two sample sizes: $n_0$ and $n_1$. We approximate the sampling distribution under $H_j$ for other sample sizes using the functions $\{\hat{L}^{_{(n)}}_{\boldsymbol{\delta}, j, r}\}_{r = 1}^m$. We use contour plots to synthesize these approximations to the sampling distributions. These plots visualize how changes to $n$ and the critical value $\gamma$ impact power and the type I error rate. 

\begin{figure}[!tb] \centering 
		\includegraphics[width = \textwidth]{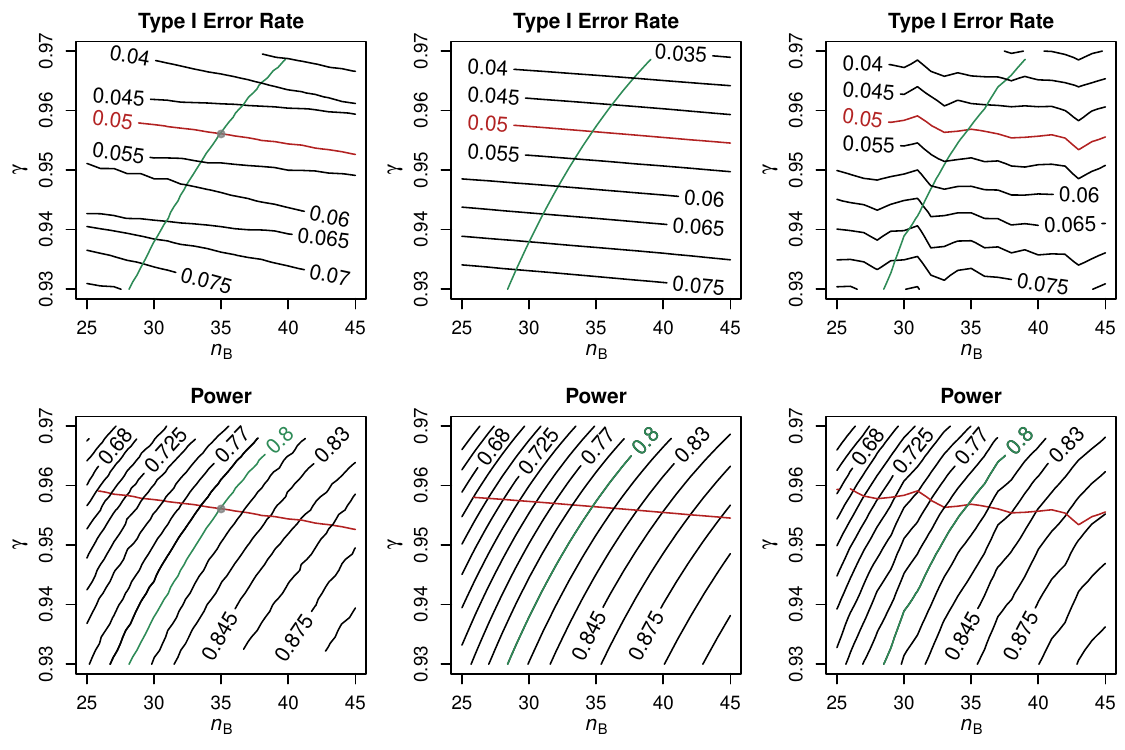} 

		\caption{\label{fig:contour} Left: Contour plots for the type I error rate and power from one sample size calculation for the semaglutide example with the optimal $(n_B, \gamma)$ combination in gray. Center: Averaged contour plots from 1000 sample size calculations. Right: Contour plots obtained by estimating sampling distributions throughout the $n_B$-space.} 
	\end{figure}

The left column of Figure \ref{fig:contour} illustrates the contour plots with respect to the type I error rate and power for the sample size calculation in Section \ref{sec:multi.ex1}. These contour plots are available with a single application of our methodology. To assist with interpretation, the green contour corresponding to power of $1 - \beta = 0.8$ and the red contour corresponding to a type I error rate of $\alpha = 0.05$ are depicted on both plots. We explored the $(n_B, \gamma)$-space in Section \ref{sec:multi.ex1}. The criteria in (\ref{eq:oc.est}) are respectively satisfied for the regions of the $(n_B, \gamma)$-space that are below the green contour and above the red contour. The optimal design for this repetition characterized by $(n_B, \gamma) = (35, 0.9561)$ is depicted by the gray point. The optimal sample size of $n_B = 35$ is the smallest $n_B \in \mathbb{Z}^+$ that is to the right of the intersection of the red and green contours. The left contour plots and the optimal sample size would differ slightly for each implementation of Algorithm \ref{alg4}, with variability decreasing as $m$ increases.

The contour plots in the left column of Figure \ref{fig:contour} are created by \emph{repurposing} logits of posterior probabilities that were computed in Algorithm \ref{alg4}. These contour plots can therefore be generated in about the same amount of time that it takes to implement Algorithm \ref{alg4} (under 4 seconds for the example in Section \ref{sec:multi.ex1}). Even if strictly controlling the type I error rate is not required, it is worthwhile to explore the sampling distributions of posterior probabilities under both $H_0$ and $H_1$; these contour plots allow practitioners to visualize the trade-off between type I and II error for various sample sizes and decision criteria without implementing extensive numerical studies. We emphasize that the contours for the type I error rate in the top left plot are not perfectly flat, which we might expect if the standard uniform approximation to the sampling distribution of posterior probabilities under $H_0$ were exact \citep{bernardo2009bayesian}. These contours may noticeably change as a function of $n$ if (i) informative priors are used, (ii) the data generation process $\Psi_0$ is nondegenerate, or (iii) moderate sample sizes are considered as also illustrated in Appendix C.1 of the supplement in the contour plots for the example from Section \ref{sec:multi.ex2}. This reinforces why we consider study design in the $(n, \gamma)$-space rather than selecting sample sizes $n$ for fixed critical values $\gamma$. However, we acknowledge that design in the $n$-space may be suitable with uninformative priors and large sample sizes; Algorithm \ref{alg4} can be simplified to accommodate these settings. 

To gain insight into how our method performs under repeated simulation, we averaged contour plots corresponding to the 1000 repetitions of the sample size calculation from Section \ref{sec:multi.ex1}. These plots are given in the center column of Figure \ref{fig:contour}, but they take 1000 times as long to generate as the left plots and are not feasible to create in practice. Based on these plots, the smallest $n_B \in \mathbb{Z}^+$ to the right of the intersection of the green and red contours is also $35$. The contour plots in the right column of Figure \ref{fig:contour} were created by simulating $m = 81920$ samples samples according to $\Psi_0$ and $\Psi_1$ for $n_B = \{25, 26, \dots, 45 \}$ following the process detailed in Algorithm \ref{alg.int}. Even for this example that leverages conjugate priors, this process takes about 80 minutes using parallelization with 72 cores. The contours in the right plots are more jagged because we obtained independent estimates for power and the type I error rate for each $n_B$ value in the plot. Nevertheless, the plots in the center and right columns are similar, which is a consequence of Theorem \ref{thm1}. The smallest $n_B \in \mathbb{Z}^+$ to the right of the intersection of the green and red contours in the right plots is $n_B = 35$. Moreover, the fact that the center and right columns of Figure \ref{fig:contour} do not differ much from the left column builds confidence in the single-application contour plots. 

\section{Discussion}\label{sec:disc}

In this paper, we developed an economical framework to design posterior analyses using operating characteristics – namely power and the type I error rate -- that determines optimal sample sizes and decision criteria. The efficiency of this framework stems from considering a proxy for the sampling distribution of posterior probabilities based on large-sample theory to justify estimating the true sampling distributions at only two sample sizes. This approach substantially reduces the number of simulation repetitions required to design posterior analyses, making them much more attractive and accessible to practitioners who want to control type I and II error. The posterior probabilities used to determine the optimal sample size and decision criteria can also be repurposed to (i) construct bootstrap confidence intervals that acknowledge simulation variability and (ii) helpfully investigate various sample sizes and decision criteria using contour plots.

Our proposed methods are broadly applicable to various two-group comparisons, including those that are more complex than our illustrative examples. Nevertheless, our methods could be extended in many aspects to accommodate more complex designs. For instance, future work could consider economical design methods that account for sequential analyses allowing for early termination or the multiple comparisons problem more generally. This extension would be nontrivial because we would need to properly account for the level of dependence in the joint distribution of the posterior probabilities across all potential analyses and estimands. We could also extend our methods to accommodate Bayesian hierarchical models for dependent data; the standard BvM theorem does not apply in those settings. Furthermore, practitioners may want to consider the operating characteristics of a posterior analysis for various data generation processes $\Psi_1$ and $\Psi_0$. While we already efficiently explore the sample size space, it would be of interest to derive analogs to Theorem 1 that enable efficient consideration of the $\Psi_1$-space and $\Psi_0$-space. 


 \section*{Supplementary Material}
These materials include a detailed description of the conditions for Theorem \ref{thm1} along with its proof and additional simulation results. The code to conduct the numerical studies in the paper is available online: \url{https://github.com/lmhagar/EconDesignPosterior}.

	\section*{Funding Acknowledgement}
	This work was supported by the Natural Sciences and Engineering Research Council of Canada (NSERC) by way of PDF and PGS-D scholarships as well as Grant RGPIN-2019-04212.
	



\bibliographystyle{chicago}


\end{document}


\newcommand{\bb}{\boldsymbol{\beta}}

	\title{An Economical Approach to Design Posterior Analyses \medskip\\
 \Large{Supplementary Material}}


\author{}

	\date{}

	\maketitle

	\maketitle

	\baselineskip=19.5pt


\appendix
\numberwithin{equation}{section}
\renewcommand{\theequation}{\thesection.\arabic{equation}}

\numberwithin{figure}{section}
\renewcommand{\thefigure}{\thesection.\arabic{figure}}

\numberwithin{table}{section}
\renewcommand{\thetable}{\thesection.\arabic{table}}

 \section{Additional Content for Theorem 1}\label{sec:proof}

	\subsection{Conditions for the Bernstein-von Mises Theorem}\label{sec:condBVM}

Theorem 1 from the main text requires that the conditions for the Bernstein-von Mises (BvM) theorem are satisfied. These conditions are described in \citet{vaart1998bvm}, starting on page 140. Conditions (B0), (B1), and (B2) concern the likelihood component of the posterior distribution for a parameter $\theta$. (B3) concerns the prior specifications for $\theta$. \citet{vaart1998bvm} uses $\theta_0$ instead of $\theta_{j,r}$ as defined in Section 3 of the main text to refer to the fixed parameter value, so we use that notation to state the conditions.

\begin{itemize}
    \item [(B0)] The observations are drawn independently and identically from a distribution $P_{\theta_0}$ for some fixed, nonrandom $\theta_0$.
    \vspace*{-2pt}
    \item [(B1)] The parametric statistical model from which the data are generated is differentiable in quadratic mean.  
    \vspace*{-18pt}
    \item [(B2)] There exists a sequence of uniformly consistent tests for testing $H_0: \theta = \theta_0$ against $H_1: \lVert \theta - \theta_0 \rVert \ge \varepsilon$ for every $\varepsilon > 0$.
    \vspace*{-2pt}
    \item [(B3)] Let the prior distribution for $\theta$ be absolutely continuous in a neighbourhood of $\theta_0$ with continuous positive density at $\theta_0$.
\end{itemize}

	\subsection{Conditions for the Asymptotic Normality of the Maximum Likelihood Estimator}\label{sec:MLE}

Theorem 1 from the main text also requires that the model $f^+(\boldsymbol{w}; \boldsymbol{\eta}^+)$ satisfies the regularity conditions for the asymptotic normality of the maximum likelihood estimator. These conditions should hold true for all $\boldsymbol{\eta}^+_{0,r} \sim \Psi_0$ and $\boldsymbol{\eta}^+_{1,r} \sim \Psi_1$. These conditions are detailed in \citet{lehmann1998theory}; they consider a family of probability distributions $\mathcal{P} = \{P_{\theta}: \theta \in \Omega \}$, where $\Omega$ is the parameter space. \citet{lehmann1998theory} use $\theta$ as the unknown parameter with true fixed value $\theta_0$, so we again state the conditions using this notation. \citet{lehmann1998theory} detail nine conditions that guarantee the asymptotic normality of the maximum likelihood estimator. We provide the following guidance on where to find more information about these conditions in their text. The first four conditions --  (R0), (R1), (R2), and (R3) -- are described on pages 443 and 444 of their text. (R4) is mentioned as part of Theorem 3.7 on page 447. (R5), (R6), and (R7) are described in Theorem 2.6 on pages 440 and 441. (R8) is mentioned in Theorem 3.10 on page 449.


\begin{itemize}
    \item [(R0)] The distributions $P_{\theta}$ of the observations are distinct.
    \vspace*{-2pt}
    \item [(R1)] The distributions $P_{\theta}$ have common support.
    \vspace*{-2pt}
    \item [(R2)] The observations are $\mathbf{X} = (X_1, ..., X_n)$, where the $X_i$ are identically and independently distributed with probability density function $f(x_i|\theta)$ with respect to a $\sigma$-finite measure $\mu$.
    \vspace*{-2pt}
    \item [(R3)] The parameter space $\Omega$ contains an open set $\omega$ of which the true parameter value $\theta_0$ is an interior point.
    \item [(R4)] For almost all $x$, $f(x|\theta)$ is differentiable with respect to $\theta$ in $\omega$, with derivative $f^{\prime}(x|\theta)$.
    \vspace*{-2pt}
    \item [(R5)] For every $x$ in the set $\{x :  f(x|\theta) > 0 \}$, the density $f(x|\theta)$ is differentiable up to order 3 with respect to $\theta$, and the third derivative is continuous in $\theta$.
    \vspace*{-2pt}
    \item [(R6)] The integral $\int f(x|\theta) d \mu(x)$ can be differentiated three times under the integral sign.
    \vspace*{-2pt}
    \item [(R7)] The Fisher information $\mathcal{I}(\theta)$ satisfies $0 < \mathcal{I}(\theta) < \infty$.
    \vspace*{-2pt}
    \item [(R8)] For any given $\theta_0 \in \Omega$, there exists a positive number $c$ and a function $M(x)$ (both of which may depend on $\theta_0$) such that $\lvert \partial^3 \text{log}f(x|\theta)/\partial \theta^3 \rvert \le M(x)$ for all $\{x :  f(x|\theta) > 0 \}$, $\theta_0 - c < \theta < \theta_0 + c$, and $\mathbb{E}[M(X)] < \infty$.
\end{itemize}

 \section{Proof of Theorem 1}\label{sec:lem1}

To prove Theorem 1, we introduce simplified notation, where $a(\delta_U, \theta_{j,r}) = a$ and $a(\delta_L, \theta_{j,r}) = c$. Moreover, we let $\Phi^{-1}(u_{j,r}) = b$, which we note is the same for both endpoints of the interval $(\delta_L, \delta_U)$. These simplifications yield the following result:
  \begin{equation}\label{eqn:logit_prob}
   \begin{split}
 & ~~~~ \text{log}\left(p^{_{(n)}}_{\boldsymbol{\delta}, j, r}\right) - \text{log}\left(1 - p^{_{(n)}}_{\boldsymbol{\delta}, j, r}\right) \\ & \approx \text{log}\left(\Phi\left(a\sqrt{n} + b\right) - \Phi\left(c\sqrt{n} + b\right)\right) - \text{log}\left(1 - \left(\Phi\left(a\sqrt{n} + b\right) - \Phi\left(c\sqrt{n} + b\right)\right)\right).
\end{split}
\end{equation}
The first derivative of (\ref{eqn:logit_prob}) with respect to $n$ is
  \begin{equation}\label{eqn:deriv1}
   \begin{split}
 & ~~~~ \dfrac{d}{dn}\left[\text{log}\left(\Phi\left(a\sqrt{n} + b\right) - \Phi\left(c\sqrt{n} + b\right)\right) - \text{log}\left(1 - \left(\Phi\left(a\sqrt{n} + b\right) - \Phi\left(c\sqrt{n} + b\right)\right)\right)\right] \\ & = \dfrac{a\phi\left(a\sqrt{n} + b\right) - c\phi\left(c\sqrt{n} + b\right)}{2\sqrt{n}(\Phi\left(a\sqrt{n} + b\right) - \Phi\left(c\sqrt{n} + b\right))} + \dfrac{a\phi\left(a\sqrt{n} + b\right) - c\phi\left(c\sqrt{n} + b\right)}{2\sqrt{n}\left(1 - \left(\Phi\left(a\sqrt{n} + b\right) - \Phi\left(c\sqrt{n} + b\right)\right)\right)}.
\end{split}
\end{equation}

We consider the limit of this derivative as $n \rightarrow \infty$ in three cases. In the first case, we consider $\theta_{j,r} \in (\delta_L, \delta_U)$ under $H_1$. In this setting, $\Phi\left(a\sqrt{n} + b\right) - \Phi\left(c\sqrt{n} + b\right) \rightarrow 1$ as $n \rightarrow \infty$. Therefore, the limit of the first fraction in (\ref{eqn:deriv1}) as $n \rightarrow \infty$ is 0. The second fraction can be written in an indeterminate form, so we consider its limiting behaviour using L'Hopital's rule. We have that
 \begin{equation}\label{eqn:Lhop}
   \begin{split}
 & ~~~~ \lim_{n \rightarrow \infty} \dfrac{\dfrac{a}{\sqrt{n}}\phi\left(a\sqrt{n} + b\right) - \dfrac{c}{\sqrt{n}}\phi\left(c\sqrt{n} + b\right)}{2(1 - (\Phi\left(a\sqrt{n} + b\right) - \Phi\left(c\sqrt{n} + b\right)))}  \\ & = \lim_{n \rightarrow \infty} \dfrac{a\left(a^2 + \dfrac{ab }{\sqrt{n}} + \dfrac{1}{n}\right)\phi\left(a\sqrt{n} + b\right) - c\left(c^2 + \dfrac{cb }{\sqrt{n}} + \dfrac{1}{n}\right)\phi\left(c\sqrt{n} + b\right)}{2\left(a\phi\left(a\sqrt{n} + b\right) - c\phi\left(c\sqrt{n} + b\right)\right)}.
\end{split}
\end{equation}

 We must consider the limiting behaviour of (\ref{eqn:Lhop}) in cases. For the points under $H_1$ where $\theta_{j,r} \in (\delta_L, \delta_U)$, $a > 0$ and $c < 0$. When $\lvert a \rvert < \lvert c \rvert$, it follows that 
 \begin{equation}\label{eqn:deriv2_ac}
   \begin{split}
 & ~~~~ \lim_{n \rightarrow \infty} \dfrac{a\left(a^2 + \dfrac{ab }{\sqrt{n}} + \dfrac{1}{n}\right)\phi\left(a\sqrt{n} + b\right) - c\left(c^2 + \dfrac{cb }{\sqrt{n}} + \dfrac{1}{n}\right)\phi\left(c\sqrt{n} + b\right)}{2\left(a\phi\left(a\sqrt{n} + b\right) - c\phi\left(c\sqrt{n} + b\right)\right)}  \\ & = \lim_{n \rightarrow \infty} \dfrac{a\left(a^2 + \dfrac{ab }{\sqrt{n}} + \dfrac{1}{n}\right) - c\left(c^2 + \dfrac{cb }{\sqrt{n}} + \dfrac{1}{n}\right)\dfrac{c\phi\left(c\sqrt{n} + b\right)}{\phi\left(a\sqrt{n} + b\right)}}{2\left(a - \dfrac{c\phi\left(c\sqrt{n} + b\right)}{\phi\left(a\sqrt{n} + b\right)}\right)} \\ & = \lim_{n \rightarrow \infty} \dfrac{a\left(a^2 + \dfrac{ab }{\sqrt{n}} + \dfrac{1}{n}\right) - c\left(c^2 + \dfrac{cb }{\sqrt{n}} + \dfrac{1}{n}\right)\text{exp}\left(\dfrac{-1}{2}\left[(c\sqrt{n} + b)^2 - (a\sqrt{n} + b)^2\right]\right)}{2\left(a - c~\text{exp}\left(\dfrac{-1}{2}\left[(c\sqrt{n} + b)^2 - (a\sqrt{n} + b)^2\right]\right)\right)} \\ & = \dfrac{a^2}{2}.
\end{split}
\end{equation}
The last step of (\ref{eqn:deriv2_ac}) follows because the limit of the exponential term in the numerator and denominator is 0 when $\lvert a \rvert < \lvert c \rvert$. When $\lvert a \rvert > \lvert c \rvert$, it follows that 
 \begin{equation}\label{eqn:deriv2_ca}
   \begin{split}
 & ~~~~ \lim_{n \rightarrow \infty} \dfrac{a\left(a^2 + \dfrac{ab }{\sqrt{n}} + \dfrac{1}{n}\right)\phi\left(a\sqrt{n} + b\right) - c\left(c^2 + \dfrac{cb }{\sqrt{n}} + \dfrac{1}{n}\right)\phi\left(c\sqrt{n} + b\right)}{2\left(a\phi\left(a\sqrt{n} + b\right) - c\phi\left(c\sqrt{n} + b\right)\right)}  \\ & = \lim_{n \rightarrow \infty} \dfrac{ a\left(a^2 + \dfrac{ab }{\sqrt{n}} + \dfrac{1}{n}\right)\text{exp}\left(\dfrac{-1}{2}\left[(a\sqrt{n} + b)^2 - (c\sqrt{n} + b)^2\right]\right) - c\left(c^2 + \dfrac{cb }{\sqrt{n}} + \dfrac{1}{n}\right)}{2\left(a~\text{exp}\left(\dfrac{-1}{2}\left[(a\sqrt{n} + b)^2 - (c\sqrt{n} + b)^2\right]\right) - c\right)} \\ & = \dfrac{c^2}{2}.
\end{split}
\end{equation}
The last step of (\ref{eqn:deriv2_ca}) follows because the limit of the exponential term in the numerator and denominator is 0 when $\lvert a \rvert > \lvert c \rvert$. When $a = -c$, the limit in (\ref{eqn:Lhop}) is $0.5\times(a^3 - c^3)/(a-c) = a^2/2 = c^2/2$. Therefore, the limit of the first derivative in (\ref{eqn:deriv1}) is $\text{min}\{ a^2, c^2 \}/2$ when $\theta_{j,r} \in (\delta_L, \delta_U)$. 

In the second case for (\ref{eqn:deriv1}), we consider points under $H_0$, where $a$ and $c$ have the same sign. When $\theta_{j,r} > \delta_U$, $c < a < 0$, and $0 < c < a$ when $\theta_{j,r} < \delta_L$. In either case, $\Phi\left(a\sqrt{n} + b\right) - \Phi\left(c\sqrt{n} + b\right) \rightarrow 0$ as $n \rightarrow \infty$. Therefore, the limit of the second fraction in (\ref{eqn:deriv1}) as $n \rightarrow \infty$ is 0. The first fraction can be written in an indeterminate form, so we consider its limiting behaviour using L'Hopital's rule. We have that
 \begin{equation}\label{eqn:Lhop2}
   \begin{split}
 & ~~~~ \lim_{n \rightarrow \infty} \dfrac{\dfrac{a}{\sqrt{n}}\phi\left(a\sqrt{n} + b\right) - \dfrac{c}{\sqrt{n}}\phi\left(c\sqrt{n} + b\right)}{2(\Phi\left(a\sqrt{n} + b\right) - \Phi\left(c\sqrt{n} + b\right))}  \\ & = \lim_{n \rightarrow \infty} -1 \times \dfrac{a\left(a^2 + \dfrac{ab }{\sqrt{n}} + \dfrac{1}{n}\right)\phi\left(a\sqrt{n} + b\right) - c\left(c^2 + \dfrac{cb }{\sqrt{n}} + \dfrac{1}{n}\right)\phi\left(c\sqrt{n} + b\right)}{2\left(a\phi\left(a\sqrt{n} + b\right) - c\phi\left(c\sqrt{n} + b\right)\right)}.
\end{split}
\end{equation}
The limit in (\ref{eqn:Lhop2}) is just $-1$ times the limit in (\ref{eqn:Lhop}). Therefore, the limit of the first derivative in (\ref{eqn:deriv1}) is $-\text{min}\{ a^2, c^2 \}/2$ when $\theta_{j,r} \notin [\delta_L, \delta_U]$.

The third and final case for (\ref{eqn:deriv1}) is when $\theta_{j,r} \in \{\delta_L, \delta_U\}$ under $H_0$. In this scenario, we conclude that the limit of both fractions in (\ref{eqn:deriv1}) is 0 without appealing to L'Hopital's rule because $\Phi\left(a\sqrt{n} + b\right) - \Phi\left(c\sqrt{n} + b\right) \rightarrow 0.5$ as $n \rightarrow \infty$. Thus, the limit of (\ref{eqn:deriv1}) as $n \rightarrow \infty$ is 0. We emphasize that $a = 0$ if $\theta_{j,r} = \delta_U$ and $c = 0$ if $\theta_{j,r} = \delta_L$. Thus, the limit of the first derivative in (\ref{eqn:deriv1}) is $\text{min}\{ a^2, c^2 \}/2 = 0$ when $\theta_{j,r} \in \{\delta_L, \delta_U\}$.

Putting the three cases together, we obtain part $(b)$ of Theorem 1: 
 \begin{equation}\label{eqn:deriv_final}
   \begin{split}
 & ~~~~ \lim_{n \rightarrow \infty} \dfrac{d}{dn}\left[\text{log}\left(\Phi\left(a\sqrt{n} + b\right) - \Phi\left(c\sqrt{n} + b\right)\right) - \text{log}\left(1 - \left(\Phi\left(a\sqrt{n} + b\right) - \Phi\left(c\sqrt{n} + b\right)\right)\right)\right] \\[1ex] & = \begin{cases}
\dfrac{\text{min}\{a^2, c^2\}}{2}, ~~~\text{if}~ \theta_{j,r} \in [\delta_L, \delta_U] \\
\dfrac{-\text{min}\{a^2, c^2\}}{2}, ~~~\text{if}~ \theta_{j,r} \notin [\delta_L, \delta_U]
\end{cases} \\[1ex] & = (0.5 - \mathbb{I}\{\theta_{j,r} \notin (\delta_L, \delta_U) \}) \times \text{min}\{a^2, c^2\}. ~~ \qed
\end{split}
\end{equation}

\section{Additional Content for the Numerical Studies}\label{sec:num.2}

\subsection{Additional Content for Example 1}\label{appd.ex1}

Here, we present additional context and numerical studies for Example 1 from Section 5.1 of the main text. For this example, the binary response $y_{i}$ that was collected for each patient $i = 1, \dots, n_A + n_B$ denotes whether the patient experienced a severe adverse event (SAE). The covariate $x_1 = \mathbb{I}(\text{Group} = A)$ is the binary treatment indicator, and $x_2$ is the patient's baseline waist weight in kilograms (kg). We use this model for illustration. 

For this example, we specified $\Psi_0$ as a degenerate process. We must therefore choose parameter values for $\boldsymbol{\beta} = (\beta_0, \beta_1, \beta_2)$ along with parameters for the normal distributions of $x_2$. We choose values for the regression parameters of $\boldsymbol{\beta}^+_{0,r} = (-2.71, \text{log}(2), 0.25)$. The value for $\beta_1 = \text{log}(2) = \text{log}(\delta_L)$ is on the boundary of the hypotheses $H_0$ and $H_1$. The semaglutide from \citet{wilding2021once} was deemed acceptable given an odds ratio (OR) of experiencing an SAE when taking the semaglutide vs.\ the placebo of 1.59. With this information in mind, an upper limit of 2 on the OR may be reasonable for a preliminary study. The choice for $\beta_0 = -2.71$ indicates that a typical patient who receives the placebo has a 6.4\% chance of experiencing an SAE, which aligns with summary statistics from \citet{wilding2021once}. The value for $\beta_2 = 0.25$ reflects patients with baseline weights equal to the $5^{\text{th}}$ and $95^{\text{th}}$ percentiles of its distribution from \citet{wilding2021once} having a 4.2\% and 9.1\% probability of experiencing an SAE, respectively. Thus, we assume there is a moderate relationship between the probability of experiencing an SAE and baseline weight. Moreover, we suppose the baseline waist circumference $x_2$ for all patients follows a $\mathcal{N}(0, 1)$ distribution after centering and scaling. In Table 1 of \citet{wilding2021once}, the mean and standard deviation of the baseline weight in kg for both groups is roughly the same, so we assume the distributions of $x_2$ are the same in both groups. The data generation process in $\Psi_1$ is such that the $\beta_1$ component of $\boldsymbol{\beta}^+_{1,r}$ is $\text{log}(1.25)$. The OR of 1.25 that characterizes this scenario is acceptable. If actually designing this study, we might want to consider power for various OR values that are less than 2.

 \begin{figure}[!b] \centering 
		\includegraphics[width = 0.8\textwidth]{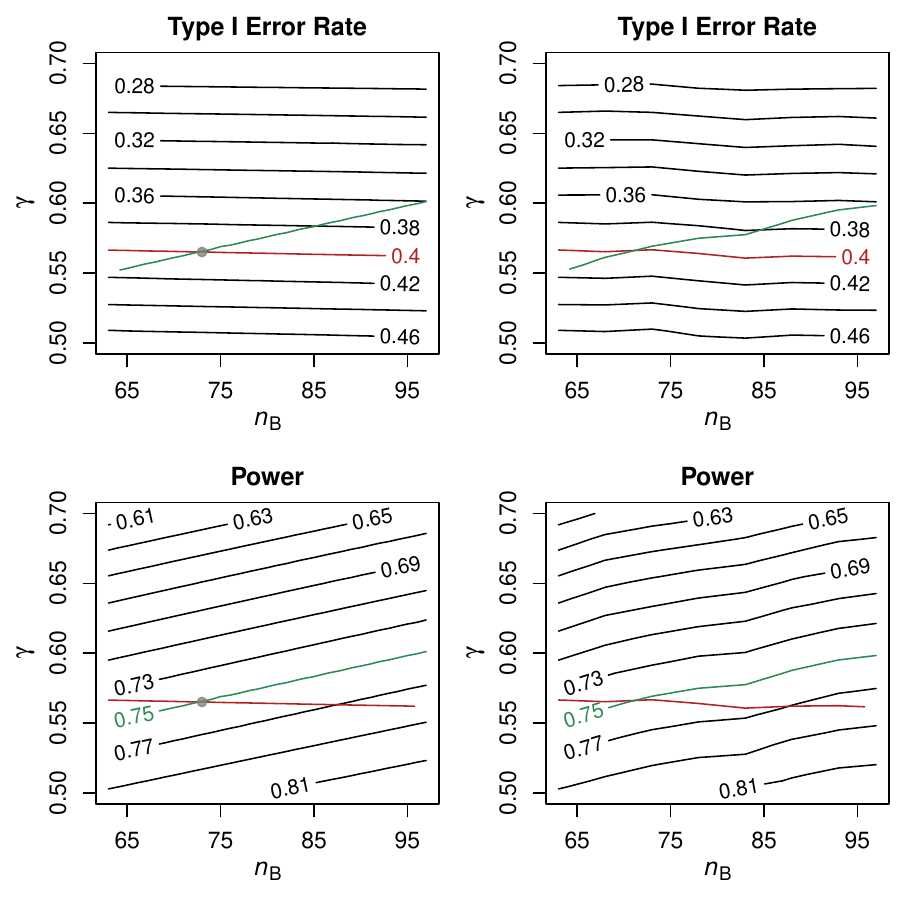} 
		\caption{\label{fig:contour.log} Left: Contour plots for the type I error rate and power from the sample size calculation for Example 1 with the optimal $(n_B, \gamma)$ combination in gray. Right: Contour plots obtained by estimating sampling distributions throughout the $n_B$-space.} 
	\end{figure}

 We followed the procedure described in Section 6 of the main text to construct contour plots using the computation from the sample size calculation from Section 5.1. These plots are given in the left column of Figure \ref{fig:contour.log}. Based on these plots, the smallest $n_B \in \mathbb{Z}^+$ to the right of the intersection of the green and red contours is $73$. The contour plots in the right column of Figure \ref{fig:contour.log} were created by simulating $m = 10^5$ samples according to $\Psi_0$ and $\Psi_1$ for $n_B = \{63, 68, \dots, 98 \}$. The smallest $n_B \in \mathbb{Z}^+$ to the right of the intersection of the green and red contours in the right plots is $n_B = 72$. The plots in the left and right columns are similar, which illustrates that using the linear approximations from Algorithm 2 to construct contour plots for this example prompts suitable performance. We emphasize that the right contour plots took four times as long to construct as the left ones because we explored the $n_B$-space without using linear approximations. Furthermore, the contours in the top plots of Figure \ref{fig:contour.log} are not perfectly flat nor perfectly calibrated such that the type I error rate when using a critical value of $\gamma$ is $1 - \alpha$.

\subsection{Additional Content for Example 2}\label{appd.ex}

We now present additional context for Example 2 from Section 5.2 of the main text. For this example, the response $y_{i}$ that was collected for each patient $i = 1, \dots, n_A + n_B$ was their percentage change in body weight over the course of the study. The covariate $x_1 = \mathbb{I}(\text{Group} = A)$ is the binary treatment indicator, and $x_2$ is the patient's baseline waist circumference in centimeters (cm). We choose to include this covariate in our linear model because it is reasonable to expect that correlation between baseline waist circumference and percentage weight loss is nonnegligible. 

For this example, we specify $\Psi_0$ as a degenerate process. We must therefore choose parameter values for $\boldsymbol{\beta} = (\beta_0, \beta_1, \beta_2)$ along with parameters for the normal distributions of $x_2$ and $\varepsilon$. Here, we choose values for the regression parameters of $\boldsymbol{\beta}^+_{0,r} = (-25.75, 5, 0.25)$. The value for $\beta_1 = 5 = \delta_L$ is on the boundary of the hypotheses $H_0$ and $H_1$. The choice for $\beta_0 = -25.75$ indicates that we expect patients in group B who are given the placebo to lose 3\% of their initial body weight on average. This assumption is reasonable since patients in both groups are given non-pharmaceutical interventions, such as counseling and diet plans. The value for $\beta_2 = 0.25$ reflects a Pearson's correlation coefficient of roughly 0.3 between baseline waist circumference in cm and percentage weight loss. 
 
 Moreover, we suppose the baseline waist circumference $x_2$ for all patients follows a $\mathcal{N}(115, 14.5^2)$ distribution to align with summary statistics from Table 1 of \citet{wilding2021once}. We suppose the error terms $\varepsilon$ follow a $\mathcal{N}(0, 10.07^2)$ distribution to reflect the confidence interval for the unadjusted treatment effect in \citet{wilding2021once} and our assumed correlation between $y$ and $x_2$. The two groups in \citet{wilding2021once} were balanced with respect to various covariates, so we assume the distributions of $x_2$ and $\varepsilon$ are the same in both groups. The data generation process in $\Psi_1$ reflects a continuum of beliefs regarding the effectiveness of the semaglutide treatment since the $\beta_1$ component of $\boldsymbol{\beta}^+_{1,r}  \overset{\text{i.i.d.}}{\sim} \mathcal{U}(9, 12)$. We have that $\beta_1 = 12$ reflects the previously demonstrated efficacy of semaglutide injections \citep{wilding2021once}, and $\beta_1 = 9$ reflects a less optimistic scenario.

\bibliographystyle{chicago}
